%% file: arxiv.tex
\documentclass[11pt]{article}

\usepackage[letterpaper,margin=1in]{geometry}
\usepackage[T1]{fontenc}
\usepackage[utf8]{inputenc}
\usepackage{lmodern}

\usepackage{amsmath}
\usepackage{amssymb}
\usepackage{graphicx}
\usepackage{booktabs}
\usepackage{siunitx}
\usepackage{url}
\graphicspath{{figures/}}

\usepackage{xcolor}
\definecolor{linknavy}{rgb}{0.0,0.2,0.55}
\usepackage{authblk}
\usepackage[colorlinks=true,linkcolor=linknavy,citecolor=linknavy,urlcolor=linknavy]{hyperref}
\hypersetup{
  pdftitle={Finite-size occupancy scaling of apparent fractal dimensions in stochastic trajectories},
  pdfauthor={Bon A. Koo, Edward Ju}}
\usepackage{xurl}

\input{generated/macros.tex}
\input{generated/macros_dna.tex}
\input{generated/macros_nondyadic.tex}

\makeatletter
\newcounter{thmcounter}[section]
\renewcommand{\thethmcounter}{\thesection.\arabic{thmcounter}}
\newenvironment{xtheorem}{\par\smallskip\refstepcounter{thmcounter}%
  \noindent\textbf{Theorem~\thethmcounter.}\itshape\space}{\par\smallskip}
\newenvironment{xdefinition}{\par\smallskip\refstepcounter{thmcounter}%
  \noindent\textbf{Definition~\thethmcounter.}\normalfont\space}{\par\smallskip}
\makeatother

\begin{document}

\title{Finite-size occupancy scaling of apparent fractal dimensions in
  stochastic trajectories}

\author[1]{Bon A.\ Koo\thanks{Corresponding author:
  \href{mailto:bkoo27@seas.upenn.edu}{bkoo27@seas.upenn.edu}}}
\author[2]{Edward Ju}
\affil[1]{Department of Computer and Information Science, University of Pennsylvania,
  Philadelphia, PA 19104, USA}
\affil[2]{Department of Computing and Mathematical Sciences, California Institute
  of Technology, 1200 E.\ California Blvd., Pasadena, CA 91125, USA}
\date{}

\maketitle

\begin{abstract}
Estimating a fractal dimension from a finite stochastic trajectory is a finite-size
scaling problem: the apparent box-counting exponent is set by an occupancy crossover
between the resolved band of scales and the finite number of sample points, and need
not equal the dimension of the limiting process. We treat this crossover as finite-size
scaling and turn it into a bias correction. A balls-in-boxes occupancy model,
$E[N(\epsilon)] = b\,[1-(1-1/b)^{N_{\mathrm{eff}}}]$ with $b=A\,(1/\epsilon)^{D_f}$ and
$N_{\mathrm{eff}}$ the number of distinct sampled sites, predicts the full box-count
curve, the finite-size saturation scale $k^\star=D_f^{-1}\log_2(N_{\mathrm{eff}}/A)$, and
a single crossover scaling function $h(\xi)$ for the normalized local slope in
$\xi=D_f(k-k^\star)$. Across walk traces (dimension $2$ for all $d\ge2$), fractional
Brownian graphs (dimension $2-H$), and L\'evy flights (range dimension $\alpha$),
the normalized local slope follows this one curve (correlation $\uniCorr$ over
$D_f\in[\uniDfLo,2]$), and the windowed bias collapses onto a single curve in
$x=k_{\mathrm{mid}}-k^\star$ that the model predicts, using $A=d$ from point spacing
and no fit to the bias data, at correlation $\modelCorr$. Fitting the model yields a corrected dimension that cuts
the walk-trace RMSE from $\traceBoxRMSE$ to $\traceOccRMSE$ and generalizes out of
class. Alternative estimators confirm the dominant bias is specific to point-sampled
box-counting over finite windows: the correlation dimension is less biased on walk
ranges, while detrended fluctuation analysis, the variogram, and Higuchi's method
recover fBm graph dimensions more accurately; the common slope-stability diagnostic
fails. We illustrate the workflow on an \emph{E.~coli} DNA walk. All figures, tables,
and in-text numbers are regenerated by released, single-seed code.
\end{abstract}


\bigskip

\section{Introduction}\label{sec:intro}
Finite systems rarely display the clean power laws of their infinite-size limits:
observables cross over between regimes set by the system's finite extent, and the
standard way to separate intrinsic behavior from finite-size artifacts is scaling
and data collapse~\cite{Privman1990}. Estimating a fractal dimension from a single
finite stochastic trajectory is such a problem. One fits a power law
$N(\epsilon)\sim\epsilon^{-D}$ to a finite set of points over a finite band of box
sizes~\cite{Falconer2003,Mandelbrot1983,Theiler1990}; the slope is read off a
\emph{finite} record---a diffusing or actively transported particle, a simulated
path, a recorded signal---and the resulting apparent exponent is a finite-size
quantity, not the dimension of the limiting process. This paper gives a
quantitative, predictive account of that finite-size crossover for stochastic
processes---random walks, anomalous (L\'evy) diffusion, and self-affine
motion---shows that it collapses onto a single scaling function, and uses the
collapse to correct the bias. The object of study is a finite-size-scaling
phenomenon of stochastic processes, not the construction of a fractal-dimension
estimator; the estimator comparison and the correction below are diagnostics of
that phenomenon.

Finite-sample bias, window dependence, and occupancy effects in box-counting are
well known. Box-counting log--log plots are curved and scatter into several
candidate slopes~\cite{Buczkowski1998,Gonzato2000}; the estimator converges slowly
and is biased low~\cite{Theiler1990,HallWood1993,GneitingSevcikova2012}; and the
finite-sample box count saturates at the number of sampled points. Early random-walk
simulations already noted the apparent box-counting dimension falling below its
theoretical value at finite length~\cite{TsurumiTakayasu1986}, without modelling the
crossover. In particular,
Kenkel~\cite{Kenkel2013} derived the same finite-sampling occupancy expectation we
use below and turned it into sample-size requirements and a small-scale cutoff,
and finite-sample corrections to dimension estimates have a long
history~\cite{Grassberger1988,BorganiMurante1994}. What remains less developed is a
finite-size-scaling treatment for stochastic trajectories: a single rescaled
coordinate locating the regression window relative to the saturation scale, a
normalized local-slope crossover tested across trajectory classes with different
limiting dimensions, and a correction validated out of class. That synthesis,
generic to stochastic trajectories rather than tied to any one application, is the
contribution here.

Two points fix the scope. First, we report a \emph{single finite-size crossover
curve across the tested stochastic-trajectory classes}---one scaling function
$h(\xi)$ onto which the data collapse once expressed in the scaling variable,
independent of microscopic model and true dimension $D_f$---not new critical
exponents or a renormalization-group universality class; the limiting dimensions
are the classical ones (Theorem~\ref{thm:taylor} and its analogues). Second, we
propose \emph{no} new estimator: box-counting, correlation dimension, detrended
fluctuation analysis (DFA), the variogram, and the Higuchi method are all standard,
and the estimator comparison is
a control that localizes the bias rather than a claim of estimator novelty.

Our organizing idea is a finite-size occupancy crossover. The empirical scaling
curve $\log N(\epsilon)$ is not a straight line: it rises at coarse scales, passes
through an intermediate band, and flattens to a plateau once boxes hold at most
one point. We model box occupancy as balls-in-boxes: if the underlying set has
true dimension $D_f$, the asymptotic covering number is $b(\epsilon)=A(1/\epsilon)^{D_f}$,
and with only $N$ sample points the expected occupied-box count is the classic
occupancy expectation~\cite{Kenkel2013}. This single model predicts the saturation
scale $k^\star=D_f^{-1}\log_2(N_{\mathrm{eff}}/A)$, a crossover scaling function for the local
slope, and the windowed bias, converting the collapse from an empirical plot into a
derived, testable prediction and a correction.

\paragraph{Contributions}
\begin{enumerate}
\item A finite-size-scaling treatment of stochastic trajectories built on the
balls-in-boxes occupancy expectation~\cite{Kenkel2013}: the trajectory saturation
scale $k^\star=D_f^{-1}\log_2(N_{\mathrm{eff}}/A)$ (equivalently $\tfrac12\log_2(N/d)$
for walk traces from the point spacing) and the full shape of the scaling curve
(Section~\ref{sec:model}, Figure~\ref{fig:loglog}).
\item A \emph{single-curve data collapse} of the normalized local slope $s(k)/D_f$
onto one crossover scaling function $h(\xi)$ in $\xi=D_f(k-k^\star)$
across the tested trajectory classes---walk traces ($D_f=2$), fBm graphs
($D_f=2-H$), and L\'evy flights ($D_f=\alpha$), spanning $D_f\in[\uniDfLo,2]$
(Section~\ref{sec:universal}, Figure~\ref{fig:universal}).
\item A \emph{scaling collapse} of the windowed box-counting bias onto a single
curve in $x=k_{\mathrm{mid}}-k^\star$ across $N$, $d$, and $\nmodels$ walk models,
quantitatively predicted by the occupancy model (Section~\ref{sec:collapse},
Figure~\ref{fig:collapse}).
\item A \emph{finite-size bias correction}: fitting the occupancy model yields a
corrected dimension that sharply reduces RMSE, validated held-out across model
classes (Section~\ref{sec:correction}, Figure~\ref{fig:correction},
Table~\ref{tab:correction}).
\item A \emph{heavy-tailed extension} to L\'evy flights (range dimension
$\alpha$), showing the crossover, collapse, and correction carry over to
anomalous superdiffusion (sections~\ref{sec:universal}--\ref{sec:correction}).
\item A broad \emph{estimator comparison} (box-counting, correlation dimension,
DFA, variogram, Higuchi) against the known values, and the finding that the
common local-slope \emph{stability} diagnostic fails---the saturated plateau is
maximally stable yet maximally biased---so the operative reliability indicator is
$x$ (sections~\ref{sec:estimators}--\ref{sec:diagnostic}).
\end{enumerate}

\paragraph{Organization} Section~\ref{sec:background} states the theory and
derives the occupancy model. Section~\ref{sec:methods} describes generators,
estimators, and statistics. Section~\ref{sec:results} reports the experiments,
building from the model to the cross-family data collapse, the correction, and an
applied example. Sections~\ref{sec:discussion}--\ref{sec:conclusion} interpret
and conclude.

\medskip
\noindent Relative to that finite-sampling occupancy work, the new content is the
finite-size-scaling theory built on top of it. Kenkel~\cite{Kenkel2013} derived the
occupancy expectation~\eqref{eq:occupancy} for a \emph{static} sampled fractal set, to
set a minimum sample size and a small-scale cutoff. We instead show that the bias of a
finite \emph{stochastic trajectory} collapses onto one curve in the rescaled window
coordinate $x=k_{\mathrm{mid}}-k^\star$ across $\nmodels$ walk models, four lengths, and
$d=2,\dots,\dmax$; that the normalized local slope follows one crossover function
$h(\xi)$ across walk traces, fBm graphs, and L\'evy flights ($D_f$ from $\uniDfLo$ to
$2$); and that inverting the model corrects the bias, validated out of class on held-out
walk models and illustrated on a measured DNA walk. Kenkel's cutoff coincides with our
$k^\star$; the collapse, the cross-class test, and the validated correction are new.
\medskip

\section{Background and the occupancy model}\label{sec:background}

\subsection{Stochastic trajectories and their limiting dimension}
\begin{xdefinition}
A \emph{simple random walk} (SRW) in $\mathbb{Z}^d$ takes unit steps along a
uniformly chosen axis; an \emph{isotropic fixed-step (Pearson)
walk}~\cite{Pearson1905} takes unit steps uniform on $S^{d-1}$; a
\emph{Gaussian (Brownian) walk} takes i.i.d.\ standard-Gaussian increments per
axis; a \emph{correlated walk} takes AR(1) Gaussian increments
$e_t=\rho\,e_{t-1}+\sqrt{1-\rho^2}\,z_t$ ($\rho>0$ persistent, $\rho<0$
anti-persistent); a \emph{L\'evy flight} takes i.i.d.\ symmetric $\alpha$-stable
increments per axis.
\end{xdefinition}

The first three walks have independent finite-variance increments and, by
Donsker's invariance principle~\cite{Donsker1951}, converge under diffusive
rescaling to $d$-dimensional Brownian motion; the correlated walks have weakly
dependent increments with summable autocovariances and positive long-run
variance, so the functional central limit theorem gives the same
limit~\cite{Billingsley1999}. Hence all $\nmodels$ share one limiting trace.

\begin{xtheorem}\label{thm:taylor}
Let $B$ be Brownian motion in $\mathbb{R}^d$. For $d\ge 2$ the trace $B[0,1]$ has
Hausdorff dimension $2$ almost surely.
\end{xtheorem}

The Hausdorff result is classical~\cite{Taylor1953}, and the Minkowski
(box-counting) dimension of the trace coincides with it~\cite{MortersPeres2010,Falconer2003}.
With the invariance principles, the limiting trace dimension of all $\nmodels$ walk
families is $2$ for every $d\ge2$, independent of $d$ and of microscopic step
details. We use two tunable benchmarks. The graph of fractional Brownian motion
$B_H$ (stationary Gaussian increments, $\operatorname{Var}B_H(t)\propto t^{2H}$)
has dimension $2-H$ almost surely~\cite{Falconer2003}. A L\'evy flight with stable
index $\alpha\in(0,2)$ has infinite-variance steps, and its range in $\mathbb{R}^d$
has Hausdorff dimension $\min(d,\alpha)$, equal to $\alpha$ in our setting
($\alpha<2\le d$)~\cite{BlumenthalGetoor1960,PruittTaylor1969};
L\'evy flights are the canonical model of anomalous (heavy-tailed, superdiffusive)
diffusion and transport~\cite{MetzlerKlafter2000}. Together these give known dimensions
spanning $D_f\in[\uniDfLo,2]$.

\subsection{Box-counting and the scaling curve}
For a bounded set $S\subset\mathbb{R}^d$ and box size $\epsilon$, let $N(\epsilon)$
be the number of $\epsilon$-grid cells meeting $S$. The box-counting (Minkowski)
dimension is $\dim_{\mathrm B}S=\lim_{\epsilon\to0}\log N(\epsilon)/\log(1/\epsilon)$
when the limit exists. A finite trajectory of $N$ steps realizes only a bounded
band of scales, so the empirical curve is not a straight line and a windowed
slope is pre-asymptotic. The object we actually box-count is the finite set of
sampled vertices---a point cloud, not the continuous limiting object---so at fine
enough $\epsilon$ the count must saturate at the number of distinct points. The
classical dimensions above ($2$, $2-H$, $\min(d,\alpha)$) are properties of the
continuous limiting paths, which the finite sample only approximates over the
resolved band; the saturation plateau analysed below is a finite-sampling effect,
not a property of those limiting paths.

\subsection{A semi-analytic occupancy model}\label{sec:model}
Work in the scale index $k$ with $\epsilon=2^{-k}$. We model the box occupancy as
balls-in-boxes. If the underlying set has true dimension $D_f$, the number of
cells a densely sampled version would occupy at scale $\epsilon$ is
\begin{equation}
b(\epsilon)=A\,(1/\epsilon)^{D_f}=A\,2^{D_f k},
\label{eq:cover}
\end{equation}
with $A$ an $O(1)$ shape/lacunarity prefactor. Distributing the $N_{\mathrm{eff}}$
distinct positions the trajectory occupies over these $b$ ``available'' cells, the
expected number of \emph{occupied} cells is the classic occupancy expectation
\begin{equation}
E[N(\epsilon)] \;=\; b\Big[1-\big(1-\tfrac1b\big)^{N_{\mathrm{eff}}}\Big]
\;\approx\;b\big(1-\exp(-N_{\mathrm{eff}}/b)\big).
\label{eq:occupancy}
\end{equation}
Here $N_{\mathrm{eff}}$ is the number of \emph{distinct} positions the trajectory
visits---its range---which sets the fine-scale plateau, and it is a property of the
trajectory rather than a fitted quantity. For non-recurrent or continuous-space
trajectories no two samples coincide, so $N_{\mathrm{eff}}$ is just the number of
sampled points (Pearson, Gaussian, and correlated walks, fBm graphs, L\'evy
flights); a recurrent lattice walk revisits sites, so $N_{\mathrm{eff}}<N$---for the
$2$-D simple random walk the range $\sim N/\log N$ puts the plateau well below $N$.
We compute $N_{\mathrm{eff}}$ directly from the simulated trajectory (the count of
distinct visited sites); when only the box-count curve is available, as for measured
data, it is estimated by the observed fine-scale plateau. Equation~\eqref{eq:occupancy}
is not new---it is Kenkel's finite-sampling expectation for a sampled fractal
set~\cite{Kenkel2013}; what is new is its use as a finite-size-scaling model for
stochastic trajectories (introduction).

Equation~\eqref{eq:occupancy} is a mean-field (independent-placement)
approximation: it treats the $N_{\mathrm{eff}}$ distinct positions as falling
independently into the $b$ cells the limiting set would occupy. Path points are
strongly correlated, but box-counting records only \emph{which} cells are occupied,
not the order of visits, and near $k^\star$ the trajectory has folded back on itself
so often that the occupied cells are spread effectively at random. Residual
correlations beyond the range renormalize the prefactor $A$, hence shift
$k^\star=D_f^{-1}\log_2(N_{\mathrm{eff}}/A)$, without changing the crossover, which
enters only through the ratio $b/N_{\mathrm{eff}}$. The shape of $h(\xi)$,
$\xi=\log_2(b/N_{\mathrm{eff}})$, is thus invariant across the lattice, Gaussian,
persistent, and anti-persistent walks despite their different local correlation
(Section~\ref{sec:universal}); the residual effects sit at the coarsest scales (the
overshoot we omit), and the strong collapse validates the approximation a posteriori
within the tested classes.
Equation~\eqref{eq:occupancy} has two regimes. When $b\ll N_{\mathrm{eff}}$ (coarse
scales) the cells are densely filled, $E[N(\epsilon)]\approx
b\propto(1/\epsilon)^{D_f}$, and the local slope tends to $D_f$. When $b\gg
N_{\mathrm{eff}}$ (fine scales) the cells are mostly empty,
$E[N(\epsilon)]\to N_{\mathrm{eff}}$, and the slope falls to $0$: the curve
saturates to a plateau. The crossover at $b(\epsilon^\star)\approx N_{\mathrm{eff}}$
defines the \emph{finite-size saturation scale}
\begin{equation}
k^\star \;=\; \frac1{D_f}\log_2\!\frac{N_{\mathrm{eff}}}{A}.
\label{eq:kstar}
\end{equation}
For a rescaled walk trace ($D_f=2$) a simple convention fixes the saturation scale
without fitting: after rescaling to $[0,1]$, a lattice walk has per-coordinate spread
$\sim\sqrt{N/d}$, so the step length relative to the diameter gives
$k^\star\approx\tfrac12\log_2(N/d)$, equivalently $A=d$ in Eq.~\eqref{eq:kstar}---the
fitting-free coordinate we adopt for the collapse. This $A=d$ is an \emph{empirical
convention}, not a first-principles constant: for Gaussian-increment walks the
unrescaled per-coordinate variance grows like $N$ rather than $N/d$, so the
$d$-dependence enters through the rescaling, and we justify it a posteriori by the
collapse. Freeing the occupancy fit instead supplies a \emph{fitted} $A$ that fixes
the curve shape, grows with $d$ (Section~\ref{sec:loglog}), and yields a saturation
scale close to this estimate---a distinct quantity from the $A=d$ convention used in
the no-regression overlay.

Two consequences make the model predictive rather than descriptive. First, the
local slope $s(k)=\mathrm{d}\log_2 E[N]/\mathrm{d}k$ depends only on the single
variable $u=b/N_{\mathrm{eff}}=2^{D_f(k-k^\star)}$:
\begin{equation}
\frac{s(k)}{D_f} \;=\; h(\xi)\;=\;1-\frac{\exp(-1/u)}{u\,[1-\exp(-1/u)]},
\qquad \xi=\log_2 u=D_f\,(k-k^\star),
\label{eq:universal}
\end{equation}
a \emph{model-independent} crossover with $h(-\infty)=1$ and
$h(+\infty)=0$. Second, the box-counting estimate over a window
$W=[k_{\min},k_{\max}]$ is the chord slope of $\log E[N]$ over $W$, so the bias
$\widehat{D}_{\mathrm B}-D_f$ is a function $F$ of the window's position relative
to $k^\star$,
\begin{equation}
x \;=\; k_{\mathrm{mid}}-k^\star \;=\; \tfrac12(k_{\min}+k_{\max})-\tfrac1{D_f}\log_2\frac{N_{\mathrm{eff}}}{A},
\label{eq:x}
\end{equation}
plus a weak dependence on window width. Section~\ref{sec:results} tests
Eqs.~\eqref{eq:occupancy}--\eqref{eq:x} directly, and Section~\ref{sec:correction}
inverts~\eqref{eq:occupancy} into a bias correction.

\section{Methods}\label{sec:methods}

\subsection{Trajectory generation}
SRW samples an axis and sign per step; Pearson normalizes Gaussian vectors to
$S^{d-1}$; the Gaussian walk accumulates standard-normal increments; persistent
($\rho=0.5$) and anti-persistent ($\rho=-0.5$) walks accumulate AR(1) increments;
L\'evy flights use the Chambers--Mallows--Stuck method for symmetric
$\alpha$-stable increments~\cite{ChambersMallowsStuck1976}. fBm graphs are
simulated exactly by Davies--Harte circulant embedding of fractional Gaussian
noise~\cite{DaviesHarte1987}.

\subsection{Box-counting, local slopes, and the occupancy fit}
We translate and isotropically rescale each trace so its largest extent fills
$[0,1]$. For $\epsilon=2^{-k}$, $k=1,\dots,\nscales$, we count distinct occupied
cells to obtain $N(\epsilon)$. The estimate $\widehat{D}_{\mathrm B}$ is the
ordinary-least-squares slope of $\log N(\epsilon)$ on $\log(1/\epsilon)$ over a
window $W$; we use OLS as the conventional box-counting slope, not as a likelihood
model (the counts across scales are dependent and heteroscedastic), and quantify
uncertainty by bootstrapping over trajectories rather than from the OLS standard
error. We define the
local slope $s(k)=\log_2 N(2^{-(k+1)})-\log_2 N(2^{-k})$ and the window stability
$S(W)=\operatorname{std}\{s(k):k\in W\}$. We \emph{fit} the occupancy
model~\eqref{eq:occupancy} to the measured $\log N(\epsilon)$ over scales with
$N(\epsilon)\ge2$: holding $D_f$ fixed yields the prefactor $A$ (hence $k^\star$),
while freeing both $D_f$ and $A$ yields the finite-size-corrected dimension of
Section~\ref{sec:correction}. To be explicit about what is and is not fitted:
(i) the windowed-bias prediction $F(x)$ overlaid on the collapse
(Figure~\ref{fig:collapse}) uses the point-spacing estimate $A=d$ (with $D_f=2$) and
no regression to the bias data; (ii) the crossover scaling function $h(\xi)$ of
Eq.~\eqref{eq:universal} is likewise a fixed analytic function, but locating each
measured curve on the $\xi$ axis uses one fitted prefactor $A$ per curve (with
$D_f$ fixed at its known value), as does the model overlay in
Figure~\ref{fig:loglog}; (iii) the bias correction fits both $D_f$ and $A$. The
box count is computed once per trajectory; all window quantities, $S(W)$, $x$, and
the fit are cheap functions of that vector. An optional GPU backend produces
counts matching the CPU path (Section~\ref{sec:repro}).

\subsection{Alternative estimators}
For cross-checks we use the Grassberger--Procaccia correlation
dimension~\cite{GrassbergerProcaccia1983} (computed on a uniform random subsample,
which suppresses serial-neighbour contamination in place of an explicit Theiler
window~\cite{Theiler1990}); for self-affine graphs, detrended fluctuation
analysis (DFA)~\cite{Peng1994,KantelhardtDFA2001} on the increments (fluctuation
exponent $=H$), the order-two variogram (log--log slope
$2H$~\cite{GneitingSevcikova2012}), and the Higuchi fractal
dimension~\cite{Higuchi1988}. The last three target the fBm graph dimension
$2-H$; since DFA and the variogram estimate the exponent $H$, we report them as the
implied graph dimension $2-\widehat H$.

\subsection{Statistics, reproducibility, and configuration}\label{sec:repro}
We report means over independent trajectories per cell and, where shown, percentile
bootstrap $95\%$ CIs. All randomness derives from a single master seed (\num{2026}),
and \texttt{code/generate\_\allowbreak paper\_\allowbreak assets.py} regenerates
every figure, table, and number in three tiers: a smoke test (\texttt{--fast}); a
default tier ($N$ up to $2^{16}$, $d$ up to $10$); and the \emph{full} tier reported
here ($N$ up to $2^{\Nmaxexp}$, $d$ up to $\dmax$, $\nscales$ scales, $50$ seeds per
cell for $N\le2^{16}$ and $20$ for larger $N$); the full tier is strictly larger than
the default in both $N$ and $d$. An optional CuPy GPU backend
(\texttt{--gpu}) counts occupied cells by a $64$-bit row hash; its counts match the
exact CPU path on the tested data (hash-collision probability $\sim10^{-7}$ per
scale at $N=10^6$), so \texttt{--full} and \texttt{--full --gpu} agree. The real-data application
(Section~\ref{sec:realdata}) is produced by
\texttt{code/analyze\_\allowbreak dna\_\allowbreak walk.py} from the bundled genome
\texttt{data/ecoli.fna.gz} (NCBI RefSeq \dnaAcc), the only non-simulated input.

\section{Results}\label{sec:results}

\subsection{The scaling curve and the occupancy model}\label{sec:loglog}
\begin{figure}[htbp]
\centering
\includegraphics[width=0.72\linewidth]{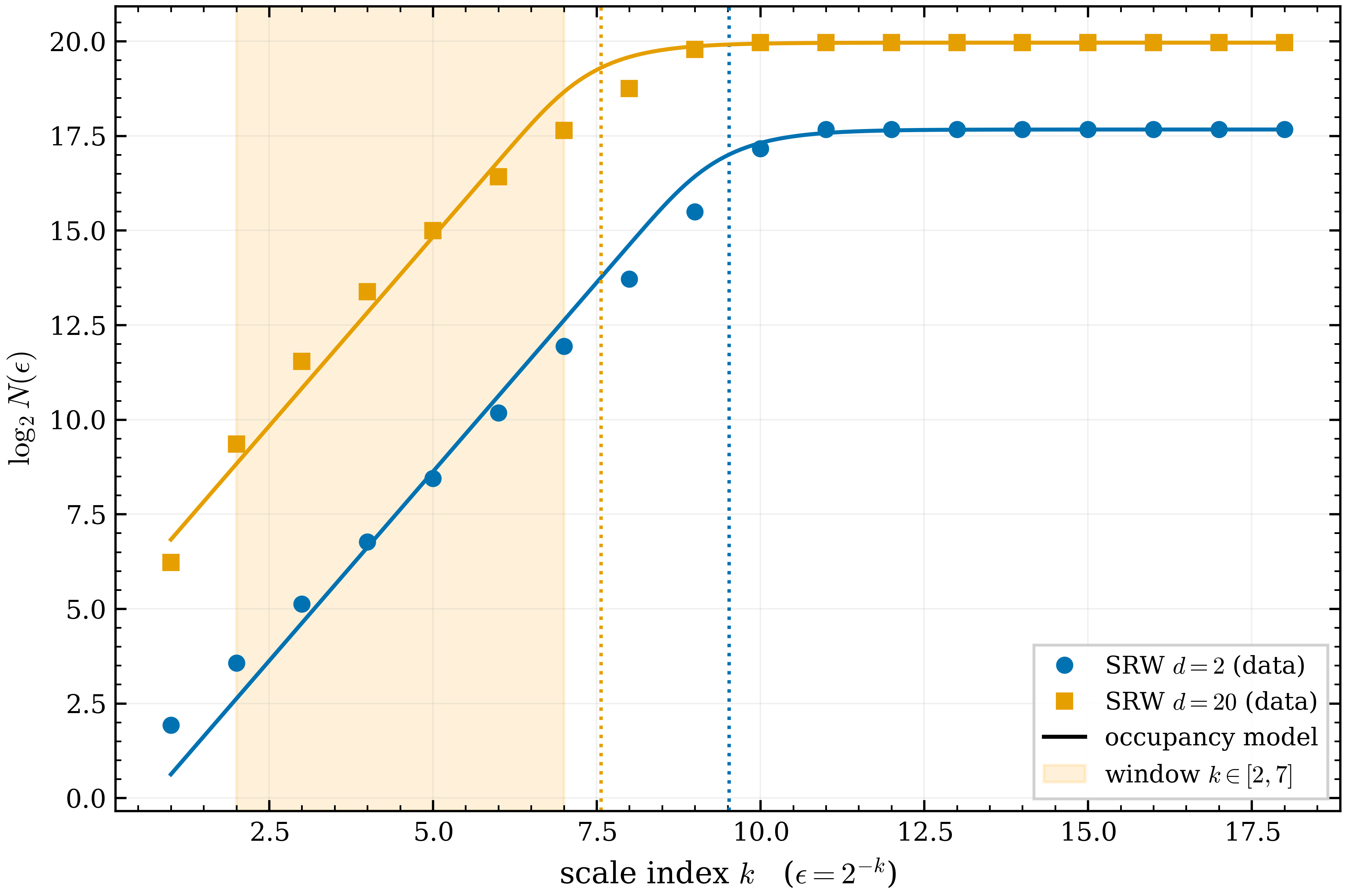}
\caption{Measured box-count scaling (markers) and the occupancy
model~\eqref{eq:occupancy} (solid, $D_f=2$, one fitted prefactor) for SRW at two
$d$ ($N=2^{\Nmaxexp}$). Dotted lines mark the fitted $k^\star$; the shaded band is
the default window.}
\label{fig:loglog}
\end{figure}

Figure~\ref{fig:loglog} overlays the occupancy model~\eqref{eq:occupancy} on the
measured box-count curves for SRW at two embedding dimensions. With $D_f=2$ fixed
and a single fitted prefactor, the model reproduces the coarse rise, the bend, and
the fine plateau across the full range of embedding dimensions. The fitted
prefactor grows with $d$ ($A=\AfitLo$ at $d=\AfitLoD$, $A=\AfitHi$ at $d=\AfitHiD$),
and the resulting saturation scale (dotted) stays close to the simple
estimate $\tfrac12\log_2(N/d)$ used for the collapse, moving to coarser $k$ as $d$
grows. The simple point-spacing coordinate is thus an accurate stand-in for the
fitted occupancy scale.

\subsection{Raw bias and its dependence on $d$, $N$, and window}\label{sec:raw}
Across $d$ at $N=2^{\Nmaxexp}$, $\widehat{D}_{\mathrm B}$ is biased low over almost
the whole grid (SRW bias $\biasTwoSRW$ at $d=2$), non-monotone in $d$, and nearly
the same across the $\nmodels$ walk models; the bias ranges from $\biasRangeLo$ to
$\biasRangeHi$, the small positive values at intermediate $d$ reflecting the
coarse-scale overshoot. Two further dependences track the same mechanism. Holding
$d=\fnDim$ and varying $N$, the estimate drifts from $\fnDlo$ at $N=2^{\Nminexp}$
toward $\fnDhi$ at $N=2^{\Nmaxexp}$; and sweeping the regression window
$[k_{\min},k_{\max}]$ for SRW at $d=2$, a single ensemble gives estimates from
$\winMin$ to $\winMax$, collapsing once $k_{\max}$ crosses $k^\star$ into the
plateau---so a number reported without its window is not interpretable. All three
dependences are what Eq.~\eqref{eq:kstar} predicts: $k^\star$ slides with
$\log_2(N/d)$ through the fixed window.

\subsection{Scaling collapse of the bias}\label{sec:collapse}
\begin{figure}[htbp]
\centering
\includegraphics[width=0.78\linewidth]{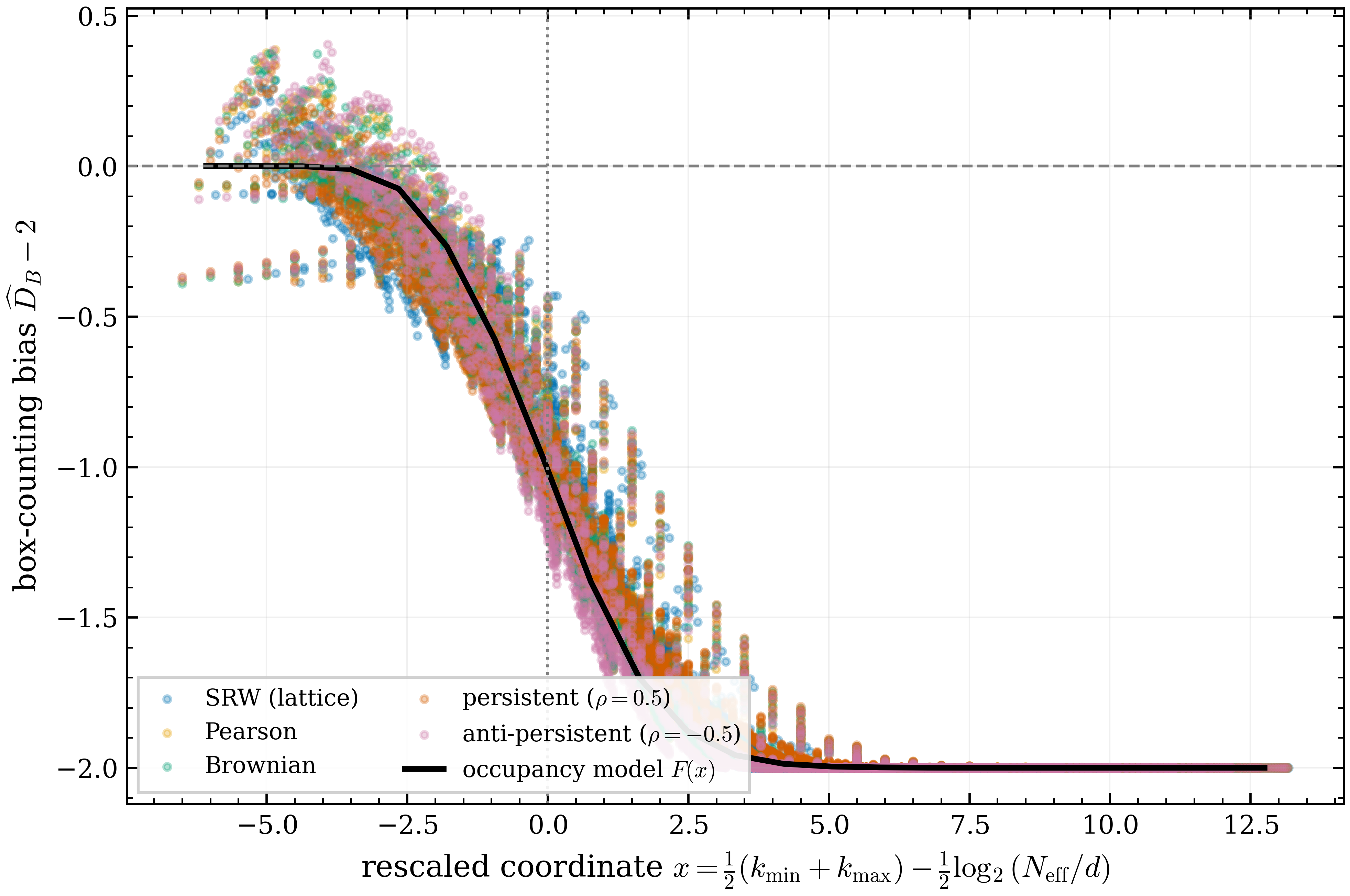}
\caption{Scaling collapse of the box-counting bias against
$x=k_{\mathrm{mid}}-\tfrac12\log_2(N_{\mathrm{eff}}/d)$ over $\nmodels$ walk models, four
sample lengths, $d=2,\dots,\dmax$, and window widths $\{\collapseWidths\}$. The black
curve is the occupancy-model prediction $F(x)$ ($D_f=2$, $A=d$ from point spacing, no
regression to the bias).}
\label{fig:collapse}
\end{figure}

For every walk model, $N$, $d$, and window of width $\{\collapseWidths\}$ we plot the
bias against $x$ of Eq.~\eqref{eq:x}, using the measured range $N_{\mathrm{eff}}$.
Figure~\ref{fig:collapse} shows the $\npts$ points collapsing onto a single curve:
rescaling reduces the spread from $\collapseRaw$ to $\collapseResid$ (a factor
$\collapseFactor$), with per-model residual at most $\collapseModelMax$, so microscopic
model class contributes little once $x$ is fixed. The recurrence correction matters
little over the tested range: the simple approximation $N_{\mathrm{eff}}\!\approx\!N$
(i.e.\ $x=k_{\mathrm{mid}}-\tfrac12\log_2(N/d)$) gives a nearly identical collapse
(residual $\collapseResidApprox$). Using the point-spacing estimate $A=d$ (with
$D_f=2$) and no regression to the bias data, the occupancy prediction tracks the
collapse, correlating with the data at $\modelCorr$; it is a derived consequence of
Eq.~\eqref{eq:occupancy}, not an empirical plotting choice. A Supplement ablation
separates this no-regression prediction from one fitted prefactor per curve and from
the fully fitted correction (Table~S1).

\subsection{Cross-family data collapse of the local slope}\label{sec:universal}
\begin{figure}[htbp]
\centering
\includegraphics[width=0.72\linewidth]{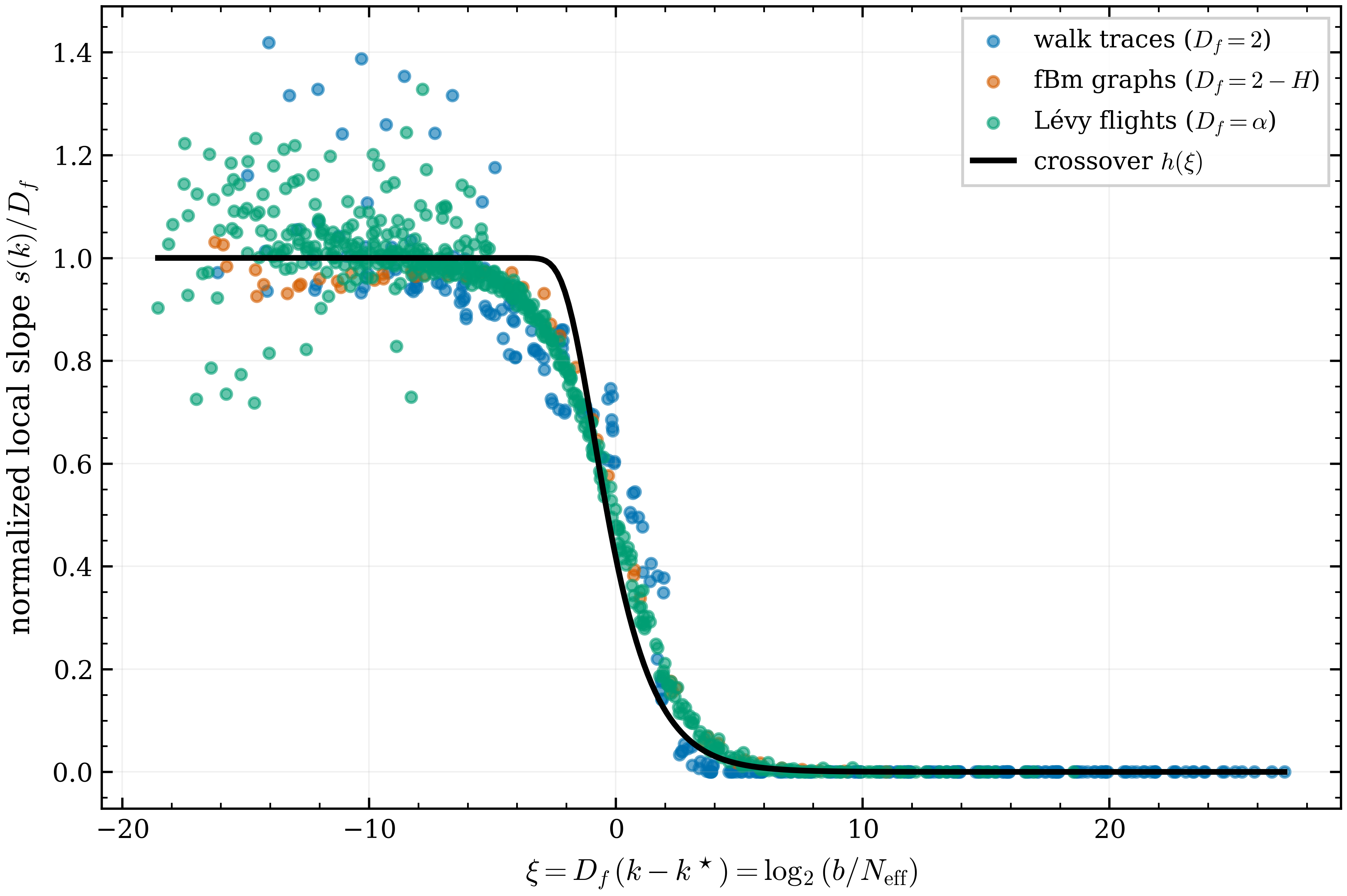}
\caption{Cross-family data collapse. Normalized local slope $s(k)/D_f$ versus the
crossover variable $\xi=D_f(k-k^\star)=\log_2(b/N_{\mathrm{eff}})$ for walk traces,
fBm graphs, and L\'evy flights, with the scaling function $h(\xi)$ of
Eq.~\eqref{eq:universal} (black). Points are the dyadic local slope of the data;
the quantitative agreement is measured against the model's matched finite-difference
slope.}
\label{fig:universal}
\end{figure}

The strongest test uses the local slope itself. Within the occupancy approximation,
Equation~\eqref{eq:universal} predicts that the \emph{normalized} local slope
$s(k)/D_f$ is a single scaling function $h(\xi)$ of the crossover variable
$\xi=D_f(k-k^\star)$, the dependence on model, $N$, $d$, and $D_f$ being absorbed
into $\xi$. Although walk ranges, fBm graphs, and L\'evy ranges are different geometric
objects, the occupancy crossover concerns the finite point-sampled covering curve and
is testable across all three.
Figure~\ref{fig:universal} overlays the
data for three families with widely differing true dimensions---walk traces
($D_f=2$), fBm graphs ($D_f=2-H$), and L\'evy flights ($D_f=\alpha$)---and the
analytic curve $h(\xi)$ of Eq.~\eqref{eq:universal}. Once each measured curve is
placed on the $\xi$ axis by its fitted saturation scale (Methods), the three families
collapse onto $h(\xi)$ with correlation $\uniCorr$ (RMSE $\uniRMSE$) over
$D_f\in[\uniDfLo,2]$. The collapse does not arise from the per-curve alignment: fixing
$A$ with no fit still gives correlation $\uniCorrNofit$, and the agreement is in
amplitude, not only rank (calibration slope $\uniCalSlope$; Supplement, Table~S1). The finite-size box-counting
crossover is therefore, within the
occupancy approximation, a single curve across the tested trajectory classes: the same
occupancy mechanism organizes the crossover for superdiffusive heavy-tailed flights,
self-affine graphs, and ordinary walk traces alike. This is an empirical
finite-size-scaling data collapse, not a new universality class.

\subsection{Finite-size bias correction}\label{sec:correction}
\begin{figure}[htbp]
\centering
\includegraphics[width=0.98\linewidth]{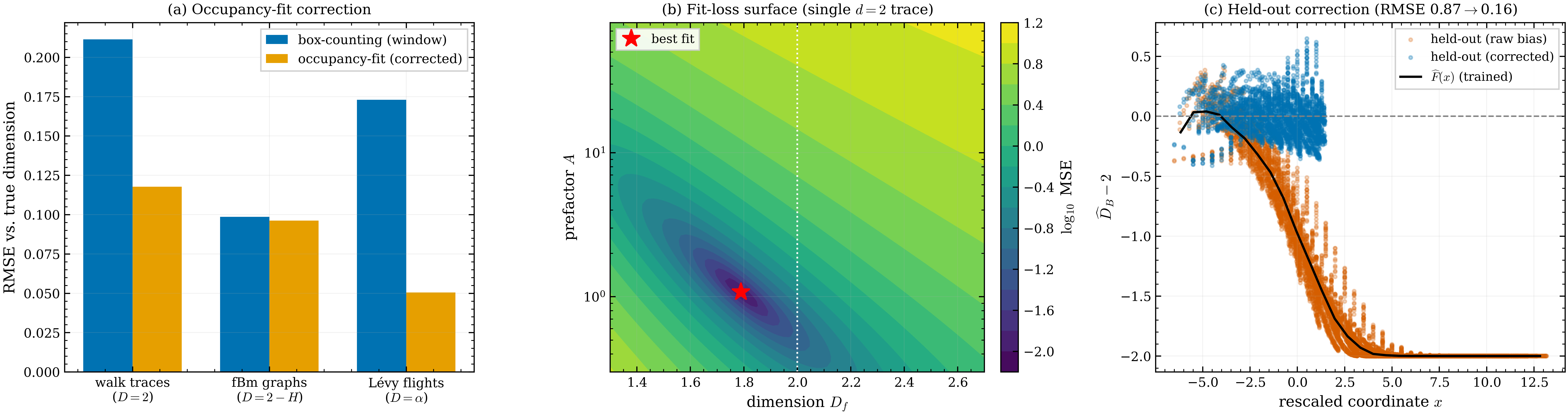}
\caption{Bias correction. (a) RMSE of the windowed box-counting estimate versus the
occupancy-fit dimension, by family. (b) Fit-loss surface ($\log_{10}$ mean-squared
error) in the occupancy parameters $(D_f,A)$ for a single $d=2$ walk trace; the single
well-defined minimum (red star; dotted line at the true $D_f=2$) shows the
two-parameter fit is identifiable from one trajectory. (c) Held-out master-curve
correction ($\widehat{F}(x)$ trained on three walk models, applied to two others).}
\label{fig:correction}
\end{figure}

If the finite-size crossover is understood, it can be undone. Finite-sample
corrections to dimension estimates are not themselves new---moment-based methods
recover the dimension of static, dilutely sampled point
sets~\cite{BorganiMurante1994,Grassberger1988,Kenkel2013}---but here both the object
and the mechanism differ: we invert the occupancy model for stochastic
\emph{trajectories} and validate the correction \emph{out of class} on held-out
trajectory models, not a bespoke estimator. Inverting
Eq.~\eqref{eq:occupancy} (fitting both $D_f$ and $A$) returns a
finite-size-corrected dimension $\widehat{D}_{\mathrm{occ}}$.
Table~\ref{tab:correction} and Figure~\ref{fig:correction}(a) compare it to
the windowed estimate against known truth. For walk traces---where box-counting is
most biased---the correction cuts the RMSE from $\traceBoxRMSE$ to $\traceOccRMSE$;
it improves the L\'evy flights ($\levyBoxRMSE\!\to\!\levyOccRMSE$) and is comparable
on fBm graphs. At the most biased cell ($d=2$ traces), box-counting reads
$\widehat{D}_{\mathrm B}=\traceBoxDtwo$ ($95\%$ bootstrap CI
$[\traceBoxDtwoLo,\traceBoxDtwoHi]$) while the occupancy fit returns
$\widehat{D}_{\mathrm{occ}}=\traceOccDtwo$ ($[\traceOccDtwoLo,\traceOccDtwoHi]$); the
two-parameter fit is well identified, its loss surface in $(D_f,A)$ showing a single
clear minimum even for one trajectory (Figure~\ref{fig:correction}(b)). On individual
tracks the fit is equally concrete: it lifts a single
$d=\appWalkD$ diffusive walk from $\appWalkBox$ to $\appWalkOcc$, and on a single
L\'evy flight ($\alpha=\appAlpha$) box-counting reads $\appBox$, the occupancy fit
$\appOcc$, and a correlation-dimension cross-check $\appCorr$. A
complementary check fits the bias curve $\widehat{F}(x)$ on three walk models and
applies it to the two held-out models (Figure~\ref{fig:correction}(c)): the
held-out RMSE falls from $\heldRawRMSE$ to $\heldCorrRMSE$, confirming the
correction transfers across model class. The plateau itself ($x\gtrsim0$) is
information-poor and not correctable; the gains are in the usable regime. We flag a fitted correction as unreliable when the
window holds fewer than four pre-plateau scales, when the bootstrap CI for
$\widehat{D}_{\mathrm{occ}}$ is wide, or when the $(D_f,A)$ loss surface lacks a single
clear minimum (Figure~\ref{fig:correction}b).

\input{generated/results_correction.tex}

\subsection{Locating the bias: process or scale window?}\label{sec:estimators}
\begin{figure}[htbp]
\centering
\includegraphics[width=0.9\linewidth]{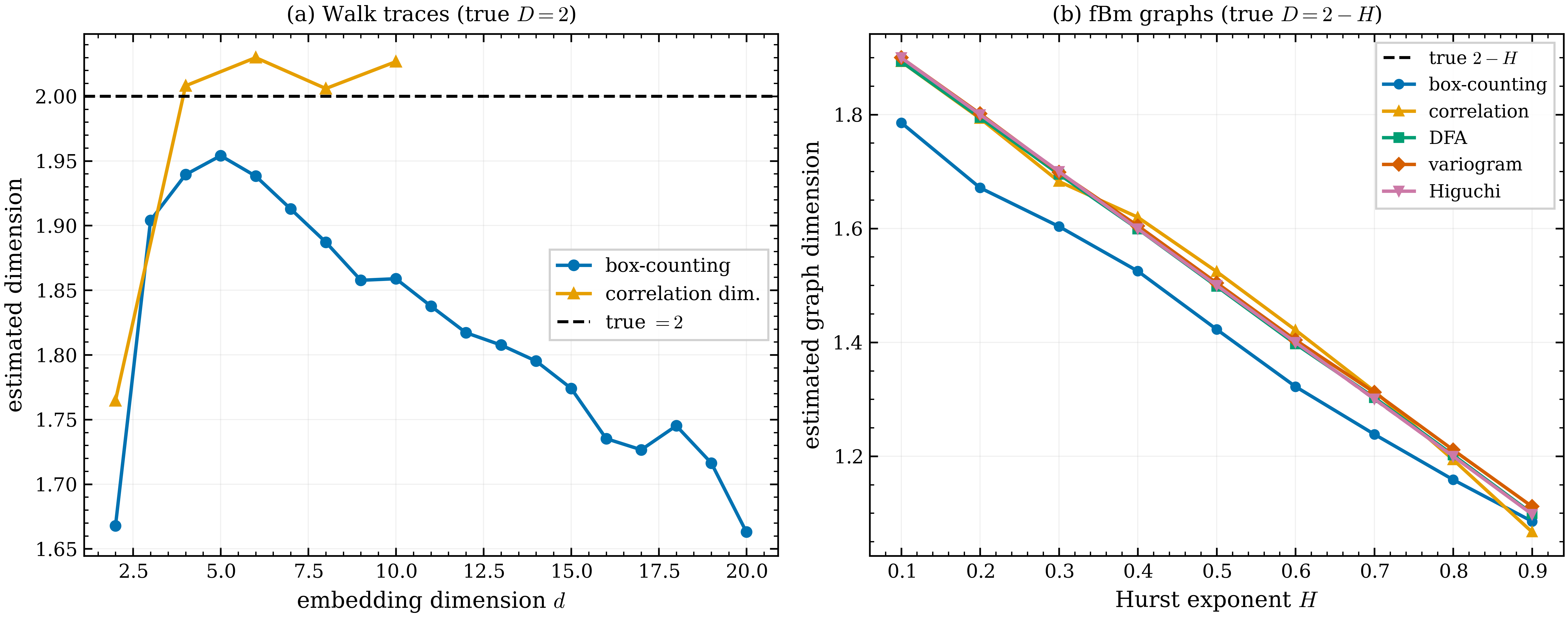}
\caption{Estimator comparison. Left: walk traces (true dimension $2$),
box-counting versus correlation dimension. Right: fBm graphs (true dimension
$2-H$), all five estimators; DFA and the variogram estimate the exponent $H$ and are
plotted as the implied graph dimension $2-\widehat H$.}
\label{fig:compare}
\end{figure}

Is the apparent-dimension bias a property of the stochastic process or of
box-counting over a finite scale window? We answer it by reading the \emph{same}
paths with estimators that weight scales differently (Figure~\ref{fig:compare}). On
walk traces the correlation dimension is far less biased than box-counting (reaching
$\corrFour$ at $d=4$, $\corrEight$ at $d=8$). On fBm graphs the structure-function
estimators are far better: DFA, variogram, and Higuchi attain RMSE $\dfaRMSE$,
$\varioRMSE$, and $\higuchiRMSE$ (correlation dimension $\corrRMSE$) against
$\boxRMSE$ for box-counting. Box-counting is worst in the rough regime: at $H=0.1$ (true
$\fbmRoughTrue$) it returns $\fbmRoughBox$ (apparent Hurst exponent
$\fbmRoughBoxH$), whereas DFA and Higuchi return $\fbmRoughDFA$ and
$\fbmRoughHiguchi$. This ordering is consistent with prior assessments of
fractal-dimension and roughness estimators, which find box-counting biased low and
structure-function (variogram/DFA-type) methods more
efficient~\cite{HallWood1993,GneitingSevcikova2012,Eke2000}. The answer to the
question above is therefore clear for these controlled classes: in the tested
synthetic trajectories the deviation is localized primarily to box-counting over its
finite scale window---a finite-size measurement effect---rather than to the
stochastic process, whose limiting dimension the scale-weighting estimators recover.
For measured data, estimator disagreement can have additional sources
(nonstationarity, multifractality, anisotropy, model misspecification), as the DNA
example below makes clear.

\subsection{Local-slope stability as a diagnostic}\label{sec:diagnostic}
A common heuristic holds that a reliable window is one over which the local slope
is stable ($S(W)$ small). We test this over every (model, $N$, $d$, window, seed)
point. Contrary to the
heuristic, the correlation of $S(W)$ with $|\widehat{D}_{\mathrm B}-2|$ is
\emph{negative} (Pearson $\diagPearson$, Spearman $\diagSpearman$; the sign is the
same for every model). The reason is structural, and follows the occupancy
picture: the most biased windows reach into the flat fine-scale plateau, where all
local slopes are near $0$ so $S(W)$ is small while the bias is large (mean
$\meanBiasPlateau$); windows below $k^\star$ are far less biased (mean
$\meanBiasSub$) but even among them $S(W)$ does not provide a useful reliability
rule (Spearman $\diagSpearmanSub$). The failure is mechanical: $S(W)$ measures the
\emph{constancy} of the local slope, not its \emph{correctness}, and the fine-scale
plateau is constant at slope $\approx0$. A window-selection rule that minimizes
$S(W)$ is therefore drawn to the most biased region, and low $S(W)$ is necessary
but not sufficient for reliability. The operative indicator is instead the window's
position relative to the saturation scale, $x$, which low $S(W)$ does not encode;
we recommend reporting both.

\subsection{Workflow demonstration: a genomic DNA walk and null controls}\label{sec:realdata}
\begin{figure}[htbp]
\centering
\includegraphics[width=0.92\linewidth]{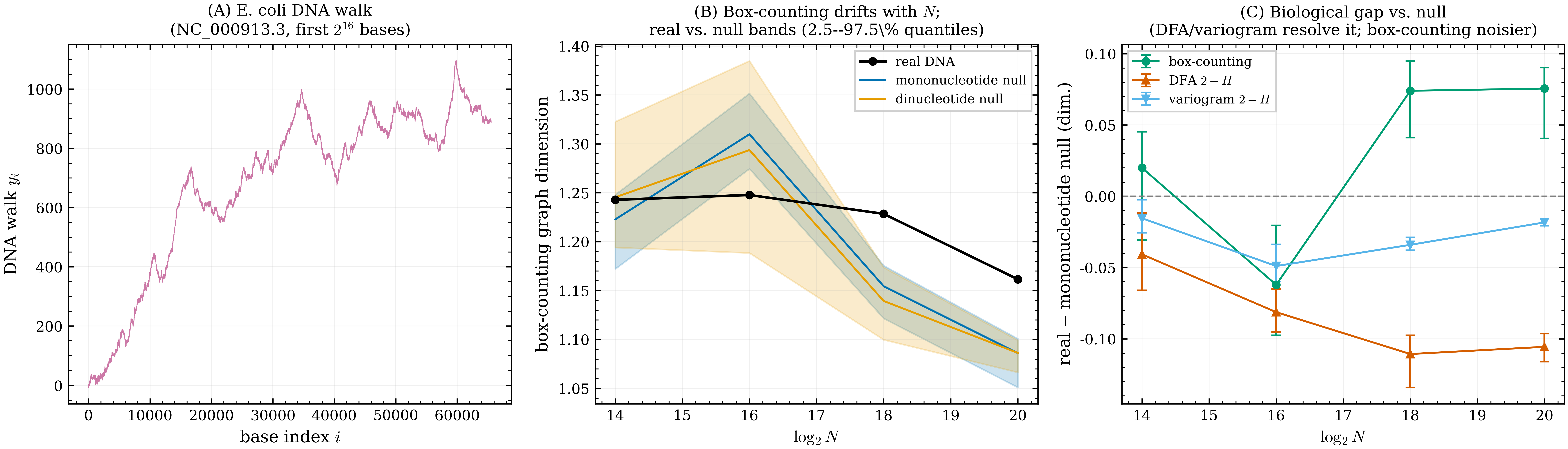}
\caption{Empirical application: the \emph{E.~coli} DNA walk (RefSeq \dnaAcc) versus
finite-size null surrogates. (A) The walk landscape. (B) Box-counting graph
dimension versus analyzed length $N$ for the real walk and the mononucleotide and
dinucleotide nulls (bands: empirical $2.5$--$97.5\%$ null quantiles); all drift with
$N$. (C) Gap between the real walk and the mononucleotide null for box-counting, DFA,
and the variogram (error bars: null $2.5$--$97.5\%$ quantiles). DFA and the variogram
resolve a stable biological
gap (the real walk's long-range correlation, $H\approx\dnaRealH$ vs.\ null
$\approx0.5$); box-counting does not.}
\label{fig:realdata}
\end{figure}

The benchmarks above use \emph{known} truth; to show the workflow on \emph{measured}
data we analyze the DNA walk~\cite{Peng1992,Peng1994} of the \emph{E.~coli} K-12 MG1655
genome (RefSeq \dnaAcc, $\dnaBases$ bases; purine $\to+1$, pyrimidine $\to-1$, cumulative
sum; Figure~\ref{fig:realdata}A). A genome's true dimension is unknown, so this is a
workflow demonstration, not a validation or biological claim---the long-range correlation
it surfaces is already established~\cite{Peng1992,Peng1994}. Because an apparent dimension
can shift from composition and short-range dependence as well as from long-range
correlation, we compare the walk to two finite-size null surrogates: a
\emph{mononucleotide} shuffle (permutes the steps; preserves composition) and a
\emph{dinucleotide} first-order Markov surrogate (matching the lag-1 purine/pyrimidine
statistics in expectation), with $\dnaNrep$ replicates per prefix
$N=2^{\dnaNlo}$--$2^{\dnaNhi}$ and an empirical $2.5$--$97.5\%$ band.

Three points follow (Figure~\ref{fig:realdata}, Table~\ref{tab:dna}). Windowed
box-counting \emph{drifts} with $N$ for the real walk and both nulls alike, with wide
bands, so a single box-counting number is not an interpretable dimension. The
structure-function estimators are stable and \emph{separate} the real walk from its
nulls: DFA gives $H\approx\dnaRealH$ versus $\dnaMonoH$ and $\dnaDinucH$ for the
mononucleotide and dinucleotide nulls (null band $[\dnaMonoHlo,\dnaMonoHhi]$), an excess
of $\dnaHexcess$. Since the dinucleotide surrogate matches composition and lag-1
dependence in expectation yet still gives $H\approx0.5$, this excess is consistent with
long-range correlation beyond composition and lag-1 dependence under these
nulls~\cite{Peng1992,Peng1994}; it does not by itself exclude higher-order Markov
structure, segmentation, or nonstationarity. The real$-$null gap
(Figure~\ref{fig:realdata}C) is resolved stably by DFA and the variogram but is small,
$N$-dependent, and sign-changing for box-counting. The finite-size cautions therefore
carry over to measured data: compare against finite-size null controls and corroborate
box-counting with a less biased estimator.

\input{generated/results_dna.tex}

\subsection{Robustness}\label{sec:robustness}
Averaging $N(\epsilon)$ over $\robustOrigins$ random grid origins shifts the estimate
by at most $\maxOffShift$, and over a fixed scale range the box-counting dimension
shifts by at most $\nondyadicShift$ across non-dyadic bases $b\in\{\nondyadicBases\}$;
these effects are intrinsic to box-counting over a finite scale band, not artifacts of
a single dyadic grid~\cite{Theiler1990}. The collapses pool the whole sweep, $N=2^{10}$
to $2^{\Nmaxexp}$ and $d=2$ to $\dmax$, so they are not tied to one scale: the
local-slope collapse (correlation $\uniCorr$), the occupancy prediction of the
windowed bias (correlation $\modelCorr$), and the residual ($\collapseResid$) are
stable across it, and the box-counting bias drifts toward the true dimension as $N$
grows (Section~\ref{sec:raw}), consistent with the finite-size effect vanishing as
$N\to\infty$.

\section{Discussion}\label{sec:discussion}
The central result is that the apparent box-counting exponent of a finite
stochastic trajectory is governed by a \emph{single finite-size occupancy
crossover} across the tested trajectory classes. The balls-in-boxes
model~\eqref{eq:occupancy} fixes the finite-size saturation scale $k^\star$,
predicts the full scaling curve (Figure~\ref{fig:loglog}), and collapses the
normalized local slope of walk traces, self-affine graphs, and L\'evy flights
onto one scaling function $h(\xi)$ over $D_f\in[\uniDfLo,2]$
(Figure~\ref{fig:universal}); the windowed bias collapses the same way against
$x=k_{\mathrm{mid}}-k^\star$, predicted at correlation $\modelCorr$ with no free
parameters (Figure~\ref{fig:collapse}). This is a statistical-mechanics
statement---a finite-size-scaling data collapse with a derived crossover variable
and scaling function~\cite{Privman1990}---not an estimator caveat.

Four consequences follow. First, microscopic details matter less than
window position: lattice, isotropic, Gaussian, and weakly correlated walks
collapse together once $x$ is fixed, and the widely reported rise of the estimate
with $d$ is a crossover, not a change in the limiting dimension, which is $2$
throughout. Second, in the tested synthetic classes the bias is a finite-size
measurement effect rather than a property of the process: estimators that weight
scales differently recover the limiting dimension on the same paths, localizing the
deviation to box-counting over its finite window (for measured data other sources
can contribute). Third, because the crossover is understood, the model is invertible:
fitting it yields a corrected dimension that cuts the trace RMSE from
$\traceBoxRMSE$ to $\traceOccRMSE$ and transfers to held-out model classes
(Figure~\ref{fig:correction}). Fourth, the intuitive slope-stability diagnostic is
misleading, because the dominant bias source---the saturated plateau---is the
flattest part of the curve; the reliable indicator is $x$.

These observations apply wherever a roughness or dimension exponent is read from
a single finite record: anomalous-diffusion and transport studies, single-trajectory
analysis, and other empirical complex systems.

\paragraph{Limitations} The collapse we document is a finite-size crossover
scaling function: a single $h(\xi)$ organizes the bias across the tested models and
true dimensions, while the limiting dimensions remain the classical values. The full tier reaches
$N=2^{\Nmaxexp}$ and $d=\dmax$ (Section~\ref{sec:robustness}); we have not explored
beyond. The plateau regime is information-poor and uncorrectable, and the model
omits the coarse-scale overshoot (a $d$-dependent effect that feeds the residual
$\collapseResid$). We verified robustness to grid origin and to non-dyadic grid
bases (Section~\ref{sec:robustness}); axis-aligned anisotropic grids are not tested.

\paragraph{When the mean-field occupancy model should fail} Equation~\eqref{eq:occupancy}
is a single-$D_f$, independent-placement approximation, and should break down where
its assumptions do: (i) \emph{multifractal} trajectories, whose local scaling is not
captured by one $D_f$; (ii) strongly \emph{anisotropic} grids or paths, where a
single saturation scale is inadequate; (iii) \emph{strong trapping or long
memory}---e.g.\ subdiffusive continuous-time random walks with diverging mean
waiting time, where ageing correlates cell occupancy beyond the prefactor
renormalization of Section~\ref{sec:model}; (iv) \emph{nonstationary} trajectories
with drift or time-varying statistics; and (v) the \emph{finite-variance crossover}
of truncated L\'evy walks, where $\alpha$-stable and Brownian regimes compete so no
single $D_f$ holds across the window. In these cases a second scaling regime or the
omitted coarse-scale overshoot would dominate and the single-curve collapse should
degrade; quantifying that is left to future work.

\section{Conclusion}\label{sec:conclusion}
Apparent fractal dimensions of finite stochastic trajectories are governed by a
single finite-size occupancy crossover across the tested classes: the balls-in-boxes
model fixes the saturation scale $k^\star$, predicts the box-count curve, and collapses
the normalized local slope of walk traces, fractional Brownian graphs, and L\'evy
flights onto one scaling function over $D_f$ from $\uniDfLo$ to $2$, while the windowed
bias collapses the same way in $x$ and inverting the model corrects it---validated out
of class and illustrated on the \emph{E.~coli} DNA walk. We therefore recommend reporting any box-counting dimension of a stochastic
path with its scale window, the window's position relative to $k^\star$, demonstrated
$N$-dependence, and a cross-check against a less biased estimator. Natural extensions
are a multifractal occupancy model and the finite-variance crossover of truncated
heavy-tailed walks.

\section*{CRediT authorship contribution statement}
\textbf{Bon A.\ Koo:} Conceptualization, Methodology, Software, Formal analysis,
Investigation, Visualization, Writing -- original draft, Writing -- review \&
editing. \textbf{Edward Ju:} Validation, Methodology, Software, Writing -- review \&
editing.

\section*{Funding}
This research did not receive any specific grant from funding agencies in the
public, commercial, or not-for-profit sectors.

\section*{Declaration of competing interest}
The authors declare no competing financial interests or personal relationships that
could have influenced the work reported in this paper.

\section*{Data availability}
All code, data, and figures are publicly available at
\url{https://github.com/Lawliet7129/fractal-occupancy-scaling}; all reported tables
and in-text numbers are regenerable from the released code under a single master seed.

\section*{Declaration of generative AI and AI-assisted technologies in the
writing process}
During the preparation of this work the authors used a generative AI assistant
(Anthropic Claude) to improve language and readability and to assist with LaTeX
formatting. After using this tool the authors reviewed and edited the content as
needed and take full responsibility for the content of the published article.


\clearpage
\appendix
\renewcommand{\thesection}{S\arabic{section}}
\renewcommand{\thesubsection}{S\arabic{section}.\arabic{subsection}}
\renewcommand{\thetable}{S\arabic{table}}
\renewcommand{\thefigure}{S\arabic{figure}}
\setcounter{section}{0}
\setcounter{table}{0}
\setcounter{figure}{0}

\begin{center}
  {\large\bfseries Supplementary Material}
\end{center}
\medskip

\noindent This appendix reports three robustness analyses referenced in the main
text: an ablation separating the no-regression collapse from fitted variants
(Section~\ref{sup:ablation}), residual diagnostics for the cross-family local-slope
collapse (Section~\ref{sup:resid}), and a Theiler-window sensitivity check for the
correlation-dimension control (Section~\ref{sup:theiler}). Every number is regenerated
by the released code from the same master seed (\num{2026}) as the main text.

\section{How much of the collapse comes from fitting?}\label{sup:ablation}
The windowed-bias collapse of the main text places each curve with the point-spacing
estimate $A=d$ and the measured range $N_{\mathrm{eff}}$, with no regression to the bias
data. Table~\ref{tab:ablation} compares this no-regression point-spacing placement
against (i) one fitted prefactor $A$ per cell ($D_f=2$ held fixed) and (ii) the fully
fitted occupancy correction ($D_f$ and $A$ both free). The no-regression collapse
residual is $\collapseResid$; allowing one fitted shift per cell tightens it only to
$\collapseResidFitA$, and the simpler $N_{\mathrm{eff}}\!\approx\!N$ approximation gives
$\collapseResidApprox$. The collapse is therefore produced by the rescaling, not by
fitting.

The same holds for the cross-family local-slope collapse. There the no-fit baseline
fixes the prefactor to a common constant ($A=1$) for every curve, with no per-curve
optimization: the measured range $N_{\mathrm{eff}}$ fixes the plateau and the placement
on the $\xi$ axis follows $k^\star=D_f^{-1}\log_2(N_{\mathrm{eff}}/A)$, while only the
fitted version optimizes $A$ per curve. This no-fit baseline still yields correlation
$\uniCorrNofit$ (RMSE $\uniRMSEnofit$) against the analytic crossover $h(\xi)$, compared
with $\uniCorr$ (RMSE $\uniRMSE$) when one prefactor is fitted per curve: the per-curve
fit improves the alignment only slightly and does not create the collapse.

\begin{table}[ht]
\centering
\caption{Ablation of the collapse against how much is fitted. The windowed-bias column
is the standard deviation of the residual bias $\widehat{D}_{\mathrm B}-2$; the
local-slope column is the correlation (and RMSE) of the normalized local slope against
the analytic $h(\xi)$. The no-regression row uses the point-spacing $A=d$ for the
windowed-bias collapse and a common constant $A=1$ for the cross-family local-slope
collapse; in both, no prefactor is optimized to the data. Each successive row fits
strictly more, yet the no-regression row already captures the collapse.}
\label{tab:ablation}
\begin{tabular}{lcc}
\toprule
prefactor placement & windowed-bias residual & local-slope corr.\ (RMSE)\\
\midrule
no per-curve regression & $\collapseResid$ & $\uniCorrNofit$ ($\uniRMSEnofit$)\\
one fitted prefactor per curve ($D_f$ fixed) & $\collapseResidFitA$ & $\uniCorr$ ($\uniRMSE$)\\
fully fitted ($D_f,A$ free) & RMSE $\traceOccRMSE$ (traces) & --\\
\bottomrule
\end{tabular}
\end{table}

\section{Residual diagnostics for the cross-family collapse}\label{sup:resid}
Correlation alone can mask amplitude bias, so we report calibration and worst-case
residuals for the cross-family local-slope collapse. Regressing the observed normalized
slope on the model prediction over $D_f\in[\uniDfLo,2]$ gives calibration slope
$\uniCalSlope$ and intercept $\uniCalInt$ (the ideal values are $1$ and $0$), with RMSE
$\uniRMSE$ and maximum absolute residual $\uniMaxResid$. The largest residuals sit at the
coarse-scale end of the crossover ($\xi\ll0$), where the mean-field occupancy model omits
the coarse-scale overshoot discussed in the main text; through the crossover and into the
fine-scale plateau the agreement is tight in both rank and amplitude.

\section{Theiler-window sensitivity of the correlation-dimension control}\label{sup:theiler}
The Grassberger--Procaccia correlation dimension~\cite{GrassbergerProcaccia1983} used as
a cross-check in the main text is computed on a uniform random subsample. An explicit
Theiler window~\cite{Theiler1990}---excluding pairs that are close in time---is standard
when a correlation dimension is estimated from a \emph{delay-embedded scalar}
time series, where temporally adjacent reconstructed points are spuriously close. For the directly observed spatial trajectories studied here, however, an explicit
Theiler window answers a slightly different question: it removes temporally adjacent
increments, which are also genuine fine-scale spatial neighbors of the sampled path. In
our controlled walk benchmark ($d=\corrTheilerD$, true dimension $2$), random subsampling
without a Theiler window recovers $\corrNoTheiler$, whereas excluding pairs within a
Theiler window of $\theilerW$ steps shifts the estimate upward to $\corrTheiler$ rather
than improving agreement with the known trace dimension. Thus the main-text
correlation-dimension comparison is not driven by serial-neighbor contamination; for the
directly sampled spatial traces considered here, random subsampling is the more
appropriate control.

\end{document}

%% file: generated/macros.tex
\newcommand{\Nmaxexp}{20}

\newcommand{\Nminexp}{10}
\newcommand{\dmax}{20}
\newcommand{\nmodels}{5}

\newcommand{\nscales}{18}

\newcommand{\collapseWidths}{4, 5, 6}
\newcommand{\npts}{22,230}

\newcommand{\biasTwoSRW}{-0.33}

\newcommand{\biasRangeLo}{-0.34}
\newcommand{\biasRangeHi}{0.02}
\newcommand{\fnDim}{4}
\newcommand{\fnDlo}{0.95}
\newcommand{\fnDhi}{1.94}
\newcommand{\winMin}{0.00}
\newcommand{\winMax}{1.78}
\newcommand{\collapseRaw}{0.67}
\newcommand{\collapseResid}{0.13}
\newcommand{\collapseFactor}{5.3}
\newcommand{\collapseModelMax}{0.13}
\newcommand{\collapseResidApprox}{0.12}
\newcommand{\collapseResidFitA}{0.10}
\newcommand{\modelCorr}{0.98}

\newcommand{\AfitLo}{0.39}
\newcommand{\AfitHi}{28.40}
\newcommand{\AfitLoD}{2}
\newcommand{\AfitHiD}{20}
\newcommand{\uniCorr}{0.98}
\newcommand{\uniRMSE}{0.08}
\newcommand{\uniDfLo}{0.6}
\newcommand{\uniMaxResid}{0.42}
\newcommand{\uniCalSlope}{0.97}
\newcommand{\uniCalInt}{+0.02}
\newcommand{\uniCorrNofit}{0.98}
\newcommand{\uniRMSEnofit}{0.10}
\newcommand{\traceBoxRMSE}{0.21}
\newcommand{\traceOccRMSE}{0.12}

\newcommand{\levyBoxRMSE}{0.173}
\newcommand{\levyOccRMSE}{0.050}
\newcommand{\heldRawRMSE}{0.87}
\newcommand{\heldCorrRMSE}{0.16}
\newcommand{\traceBoxDtwo}{1.67}
\newcommand{\traceBoxDtwoLo}{1.64}
\newcommand{\traceBoxDtwoHi}{1.69}
\newcommand{\traceOccDtwo}{1.72}
\newcommand{\traceOccDtwoLo}{1.70}
\newcommand{\traceOccDtwoHi}{1.74}
\newcommand{\diagPearson}{-0.68}
\newcommand{\diagSpearman}{-0.90}

\newcommand{\diagSpearmanSub}{0.03}
\newcommand{\meanBiasPlateau}{1.67}
\newcommand{\meanBiasSub}{0.18}

\newcommand{\boxRMSE}{0.099}
\newcommand{\dfaRMSE}{0.010}
\newcommand{\varioRMSE}{0.024}
\newcommand{\higuchiRMSE}{0.006}
\newcommand{\corrRMSE}{0.073}
\newcommand{\theilerW}{100}
\newcommand{\corrNoTheiler}{2.00}
\newcommand{\corrTheiler}{3.09}
\newcommand{\corrTheilerD}{4}
\newcommand{\fbmRoughTrue}{1.90}
\newcommand{\fbmRoughBox}{1.79}
\newcommand{\fbmRoughBoxH}{0.21}
\newcommand{\fbmRoughDFA}{1.89}
\newcommand{\fbmRoughHiguchi}{1.90}
\newcommand{\corrFour}{2.01}
\newcommand{\corrEight}{2.01}

\newcommand{\appAlpha}{1.5}
\newcommand{\appBox}{1.35}
\newcommand{\appOcc}{1.42}
\newcommand{\appCorr}{1.52}
\newcommand{\appWalkD}{6}
\newcommand{\appWalkBox}{1.90}
\newcommand{\appWalkOcc}{1.95}
\newcommand{\robustOrigins}{16}
\newcommand{\maxOffShift}{0.06}

%% file: generated/macros_dna.tex
\newcommand{\dnaAcc}{NC\_000913.3}
\newcommand{\dnaBases}{4,641,652}
\newcommand{\dnaNrep}{20}
\newcommand{\dnaNlo}{14}
\newcommand{\dnaNhi}{20}

\newcommand{\dnaRealH}{0.61}

\newcommand{\dnaMonoH}{0.50}
\newcommand{\dnaMonoHlo}{0.49}
\newcommand{\dnaMonoHhi}{0.51}
\newcommand{\dnaDinucH}{0.50}

\newcommand{\dnaHexcess}{0.11}

%% file: generated/macros_nondyadic.tex
\newcommand{\nondyadicShift}{0.04}
\newcommand{\nondyadicBases}{1.5, 2.0, 2.5, 3.0}

%% file: generated/results_correction.tex
\begin{table}[t]
\centering
\caption{Finite-size bias correction. For each model family we compare the windowed box-counting estimate (window $k\in[2,7]$) with the occupancy-fit dimension (the $D_f$ of the best-fit occupancy curve), reporting mean bias and RMSE against the known dimension. Traces use $N=2^{20}$ over $d=2,\dots,20$; fBm over $H$; L\'evy over $\alpha$. The occupancy-fit correction reduces the RMSE for the walk traces and L\'evy flights, where the finite-window bias is largest, and is essentially neutral on fBm graphs.}
\label{tab:correction}
\begin{tabular}{lcccc}
\toprule
family & box bias & box RMSE & occ.-fit bias & occ.-fit RMSE\\
\midrule
walk traces ($D=2$) & $-0.182$ & $0.211$ & $-0.073$ & $0.118$\\
fBm graphs ($D=2-H$) & $-0.076$ & $0.099$ & $-0.080$ & $0.096$\\
L\'evy flights ($D=\alpha$) & $-0.030$ & $0.173$ & $-0.012$ & $0.050$\\
\bottomrule
\end{tabular}
\end{table}

%% file: generated/results_dna.tex
\begin{table}[t]
\centering
\caption{Real \emph{E. coli} DNA walk versus finite-size null surrogates at $N=2^{20}$ (nulls: mean over 20 replicates). The mononucleotide shuffle preserves composition; the dinucleotide (first-order Markov) surrogate matches the lag-1 statistics in expectation. Box-counting drifts with $N$ and cannot be read as a fixed dimension; the structure-function estimators are stable. The real walk's Hurst exponent exceeds both null bands, consistent with long-range correlation beyond composition and lag-1 dependence under these null models; the true dimension is unknown, so this is an empirical comparison, not a validation against known truth.}
\label{tab:dna}
\small
\begin{tabular}{lccccccc}
\toprule
sequence & box & occ.-fit & DFA & variogram & Higuchi & $H$ & box drift\\
\midrule
real DNA & $1.16$ & $1.21$ & $1.39$ & $1.34$ & $1.50$ & $0.61$ & $-0.08$\\
mononucleotide null & $1.09$ & $1.21$ & $1.50$ & $1.36$ & $1.52$ & $0.50$ & $-0.14$\\
dinucleotide null & $1.09$ & $1.20$ & $1.50$ & $1.36$ & $1.52$ & $0.50$ & $-0.16$\\
\bottomrule
\end{tabular}
\end{table}